**Petrophysical characterization of fractured limestone from Beauce Aquifer vadose zone (O-ZNS Observatory, France)**

Abdoul Nasser Yacouba[1,2*], Céline Mallet[1*], Jacques Deparis[2], Philippe Leroy[2], Damien Jougnot[4], Mohamed Azaroual[1,2]

[1] *Univ. d'Orléans, CNRS, BRGM, ISTO, UMR 7327, F-45071, Orléans, France*
[2] *French geological survey, BRGM, F-45100, Orleans, France*
[3] *Univ. Orleans, Univ. Tours, INSA-CVL, LaMé EA7494, 8 rue Léonard de Vinci, F-45072 Orleans, France*
[4] *Sorbonne Université, CNRS, EPHE, UMR 7619 METIS, F-75005 Paris, France*

**Introduction**

In recent years, water needs increased tremendously, driven by climate change consequences and world population growth. In this context, it is of critical importance to study groundwater behaviour and especially the vadose zone which play a crucial role in groundwater recharge and propagation of pollutant (Stephens, 1996).
Most of the time unsaturated, this vadose zone, located between soil and water table, is characterized by multi-scale heterogeneities especially when dealing with lacustrine limestones (Tucker & Wright, 2009). They are defined by the overall pore structures (e.g., size, number, distribution, tortuosity, constrictively, presence of fractures and connections) and are enhanced by diagenetic processes and weathering in complex environments like the vadose zone. This leads to uncertainties for reservoir properties prediction using geophysical methods which impacts reservoir models for flow simulations. Among these methods, acoustics and electrical methods are well suited for reservoir characterization as there is a strong link between pore structures and both acoustics and electrical properties (Mavko et al., 2009).
In this study, we combined microstructure description and petrophysical analysis in order to model and predict reservoir properties based on different limestones facies in complex setting and to infer the influence of weathering/fracturing on both acoustics and electrical properties.

**Geological setting and sampling**

The Beauce aquifer extends over nearly 9,000 km$^2$ and constitutes the largest fresh water reservoir of France and the second largest in Europe. To understand its dynamic and characterize its vadose zone, the O-ZNS (standing for "Observatoire des transferts dans la Zone Non Saturée", in French) is set-up at Villamblain (Loiret – France). It is composed by a main well of 20 m-depth and 4 m-diameter and surrounded by 8 cored boreholes (Figure 1a –Mallet et al., 2022).
The vadose zone of Beauce aquifer is an extremely complex porous and fractured medium mainly composed by lacustrine limestones known as Beauce Limestones (Ménillet & Edwards, 2000). The formation has mainly experienced several periods of erosion and weathering leading to important karstification. In terms of lithofacies, the Beauce Limestone is very heterogeneous with great lateral and vertical variabilities making the distinction between lithofacies very complex. At the study site, the formation is located between 2 and 20 m deep and the current water table level varies between 17 and 20 m deep depending on the season.
Based on the analysis of well logs (acquired in boreholes), we identified two layers (Figure 1b). A first one corresponding to the soil and the unconsolidated limestones and a second one corresponding to the consolidated limestone on which we focus this study. This second layer shows variations in the presence of fractures, and more or less massive limestone structures. In order to investigate strongly different rock structures, we further subdivided this consolidated part in different facies (Figure 1b).

**Methods**

A total of 16 samples from these facies (named A, B, C and C') were cored and characterized by their porosity, permeability, acoustic velocities, complex electrical properties and microstructure analysis.

Effective porosity, bulk densities and grain density were measured based on hydrostatic weighing method. It is also measured with Hg-injection together with pore space properties (pore size distribution and average pore diameter). Water permeability was measured with a triaxial cell in steady-flow condition.

The ultrasound acoustic measurement was carried out in dry and saturated (water) conditions with longitudinal (P waves) and transverse (S waves) piezoelectric transducers (PZT) at 0.5 MHz.

The complex conductivity or Spectral Induced Polarization (SIP) measurements (Binley & Slater, 2020) were carried out with the SIP-LAB-IV apparatus using a 4-electrode setup where the potential (measuring) electrodes are non-polarizable Ag/AgCl electrocardiograms (ECG) and the current (injecting) electrodes are made of carbon films (Jougnot et al., 2010).

The structure and the microstructure of the samples were accessed based on photogrammetry (2D/3D images of the external structure) and X-ray microtomography (2D/3D images of both external and internal structures).

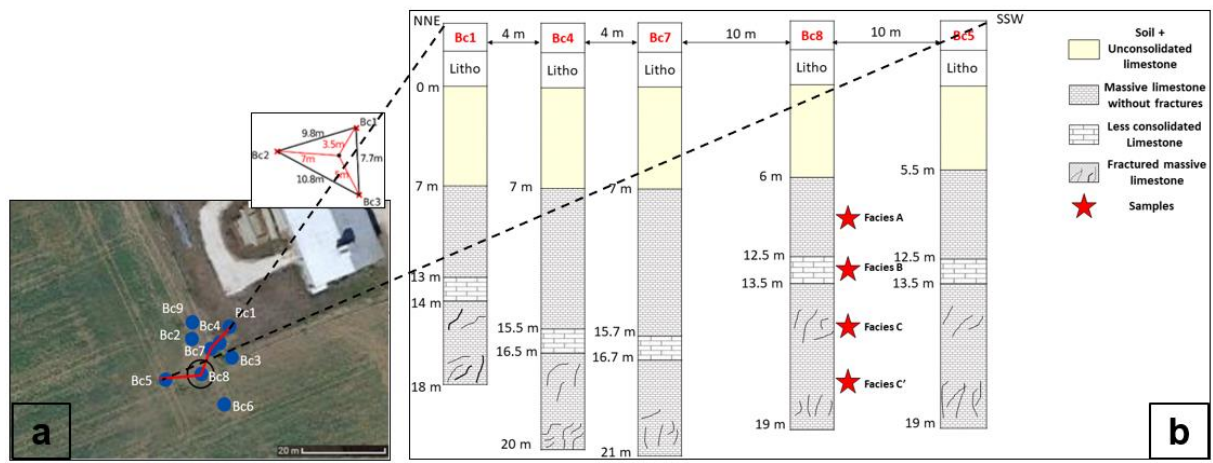

*Figure 1. a) Basemap of O-ZNS at Villamblaim (Loiret, France) with the location of the main well (black circle) and the different boreholes (filled blue circles). b) Well-well correlation sketch based on well logs (e.g., resistivity, neutron, density) acquired in the boreholes following the red line in a).*

**Heterogeneous petrophysical properties**

Reservoir properties are globally scattered and heterogenous for facies A, B and C. Only facies C', located at the water table level shows homogenous values. This facies has the highest effective porosity (11.6%) whereas facies C shows the lowest values (between 3.8 and 5.4%), facies A and B are close (6.2 and 7.4%). Conversely, permeability values which are globally low show a different trend. Indeed, facies C' shows the highest porosity but the lowest permeability (0.4 mD). Facies A and C on their side are characterized by relatively high permeabilities (between 1.7 and 7.7 mD). This is in good agreement with the pore size (average pore diameter) where we noticed that facies A and C have the largest average pore diameters (between 0.015 to 0.037 µm), while B and C' have the smallest ones (0.008 µm in average).

Acoustic velocity results show globally coherent values. A detailed analysis shows that facies C' has the lowest P- velocities both in dry and water saturated conditions with the lowest dispersion. The other facies are much more disperse, especially facies C with the highest velocities in both conditions. Facies A and B have very similar P-wave velocities in average both in dry and saturated conditions, with equivalent dispersion for both. We observe that saturating the pore with water tends to homogenize and increase $V_P$.

Regarding electrical properties, factor formation (defined as the ratio between water and rock conductivities (Archie, 1942)) and surface conductivity were obtained using a multi-salinity approach. Both formation factor (FF) and surface conductivity variations are not pronounced. Indeed, FF and surface conductivity range between 86 and 378 and 0.0008 and 0.022 S/m, respectively where Facies C' shows the lowest values.

In overall, based on reservoir and physical properties analysis, facies C' stand out as the most homogenous and consistent facies whereas facies C shows the most heterogenous values. Facies A and B are close in terms of physical properties but slightly different in terms of reservoir properties.

**Microstructure characterization**

The Beauce Limestones rock fabrics show that it is globally mud supported limestones highlighted by a well lithified or cemented micritic matrices. Some samples present more than one matrix (with different origins) and clasts (with various size) suggesting brecciated limestones. In addition, the different samples are characterized by heterogeneous features such as fractures, cracks or vugs, which make it possible to differentiate or discriminate the different facies (figure 2). Indeed, in terms of microstructure we have micropores (essentially found in the matrices) assessed by Hg injection method (average pore diameter) and macropores materialized by vugs and fractures.

The obtained CT-scan images resolution (a voxel is approximatively 20 µm) is not high enough to observe micropores, however macropores can be well characterised based on segmented images. We observe that facies A and C have the highest proportion of macropores compared to facies B and C'. Facies A and C are vuggy and fractured explaining their relatively high permeability values whereas microporous facies B and C' show low permeability values.

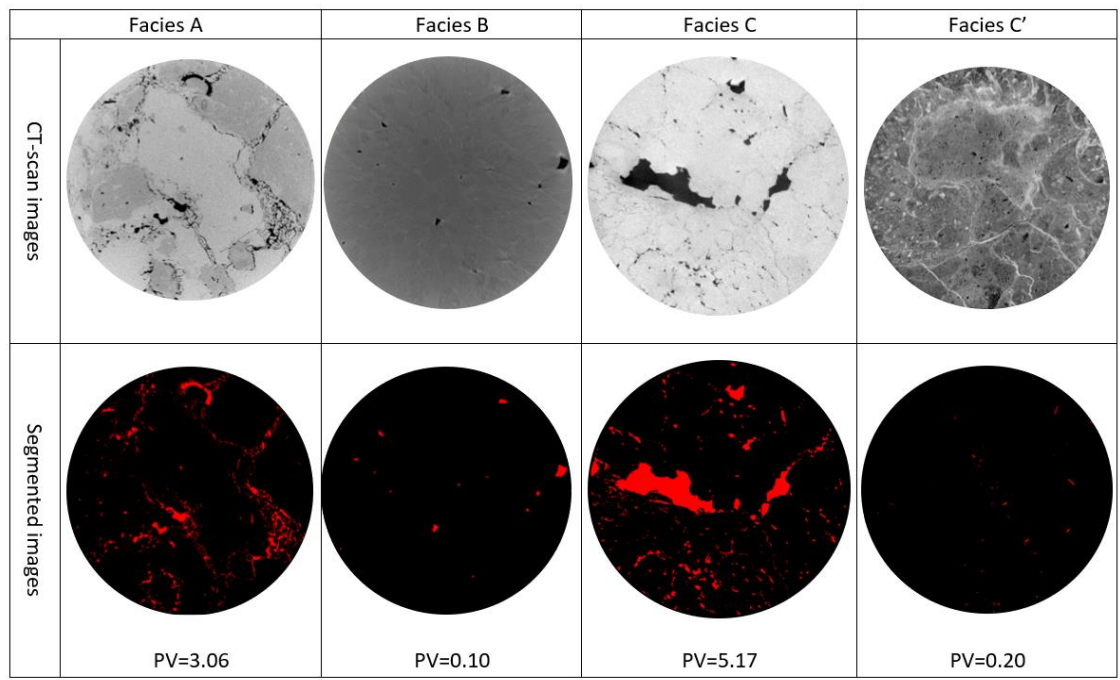

*Figure 2: X-ray scans images of the different facies obtained. The top images are slices processed in 8-bit presenting the original texture of each facies and the bottom images are the segmented images where the red colour highlights the macro-pore volume (PV in %).*

**Pore type/structure control on acoustic and electrical properties**

Figure 3 shows relationship between effective porosity and both acoustic and electrical properties. For acoustic properties (Figure 3a), the primary control of carbonate reservoirs is the porosity as it has shown by many studies. This control results in a decrease in velocities (here water saturated $V_P$) when the effective porosity increases, even though some discrepancies linked to the pore structure are observed. Indeed, the presence of macropores (e.g., Facies C) tends to scatter values and minimize the control of the effective porosity. By using empirical models, we showed that velocities predicted by Raymar-Hunt-Gardner's (RHG) model greatly overestimate the velocities while Willie Time Average (WTA) ones are closer, especially for samples with high density of microporosity (e.g., Facies C').

For electrical properties, the control of pore structure/type has been highlighted by some studies through the literature. Figure 3b shows the relationship between FF and effective porosity where we see a poor correlation considering all the samples. This relationship becomes better and compatible with Archie's law by removing facies C. Globally, the cementation factor of the Beauce limestone varies between 1.5 to 2.4 which is globally consistent for carbonates (Mavko et al., 2009). However, the microstructure, in particular the presence of vugs and fractures have an influence on the cementation exponent (facies C) which is often considered as a solely lithological parameter. Although facies A and C appear to have similar microstructures, they exhibit a different cementation factor, which shows that in certain cases, the lithological (filled or open fractures) control might be greater.

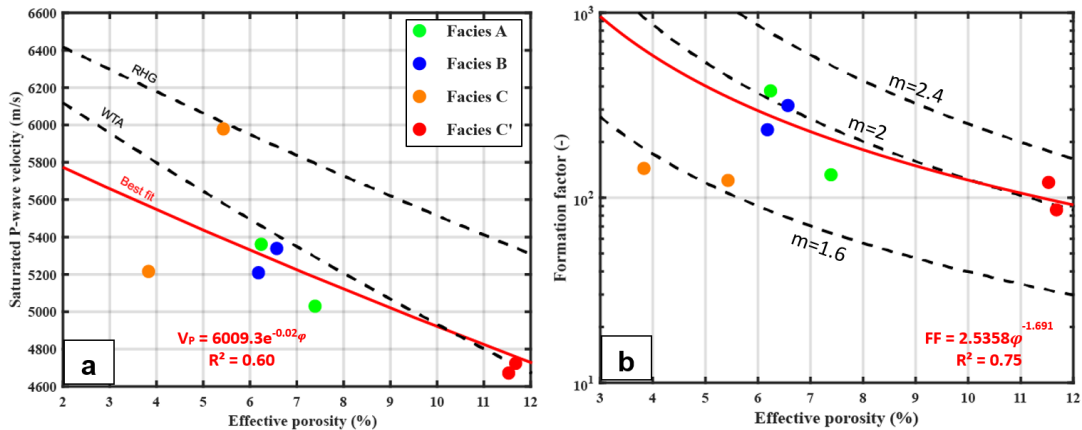

*Figure 3. a) Relationship between saturated $V_P$ and effective porosity with two empirical models in dashed lines (see Mavko et al., (2009) for equations). The red line corresponds the best fit regression line. b) Relationship between FF and effective porosity where the dashed lines are the Archie's models with different cementation exponent and the red line is the regression line (following Windsauer et al. (1952) model) without facies C samples.*

**Conclusions**

Based on a multi-method approach, we demonstrated the influence of rock structure on reservoir properties prediction and modelling in the complex vadose zone. Petrophysical and microstructure characterization have highlighted two main facies (microporous and homogenous facies and macroporous and heterogenous facies) which can be used to improve reservoir and flow models. However, further development is needed in order to quantify macropores and their link with weathering and to assess permeability models using electrical properties.